\documentclass[prl,twocolumn,superscriptaddress,amsmath,amssymb]{revtex4}
\usepackage{graphicx}
\usepackage{bm}

\makeatletter

\providecommand{\tabularnewline}{\\}

\newcommand {\Fig}[1] {Figure~\ref{#1}}
\newcommand {\fig}[1] {Figure~\ref{#1}}   

\newcommand {\eqn}[1] {Equation~(\ref{#1})}
\newcommand {\tab}[1] {Table~\ref{#1}} 

\newcommand{\beq}{\begin{equation}}
\newcommand{\eeq}{\end{equation}}

\newcommand{\erscn}{Er\-Sc$_{2}$\-N}

\newcommand{\erscnC}{Er\-Sc$_{2}$\-N\-@C$_{80}$}

\newcommand{\ernC}{Er$_{3}$\-N\-@C$_{80}$}

\newcommand{\gs}{$^{4}I_{15/2}$}

\newcommand{\ceighty}{C$_{80}$}

\newcommand{\beqa}{\begin{eqnarray}}
\newcommand{\eeqa}{\end{eqnarray}}

\newcommand{\sitea}{\emph{Conf~1}}
\newcommand{\siteb}{\emph{Conf~2}}
\newcommand{\sitei}{\emph{Conf~i}}

\begin{document}

\title{Electron paramagnetic resonance study of \erscnC}

\author{Rizvi Rahman\footnote{Both authors contributed equally to this work}}
\email{rizvi.rahman@materials.ox.ac.uk}
\affiliation{Department of Materials, Oxford University, Oxford OX1 3PH, UK}

\author{Archana Tiwari$^*$}
\affiliation{Department of Physical Sciences, Sikkim University, Gangtok 737102, India}

\author{G\'{e}raldine Dantelle}
\affiliation{Laboratoire de Physique de la Matière Condensée – CNRS UMR 7643 – Ecole Polytechnique, Route de Saclay, 91128 Palaiseau Cedex, France}

\author{John~J.~L.~Morton}
\email{john.morton@materials.ox.ac.uk}
\affiliation{Department of Materials, Oxford University, Oxford OX1 3PH, UK}
\affiliation{CAESR, Clarendon Laboratory, Dept. of Physics, Oxford University, Oxford OX1 
3PU, UK}

\author{Kyriakos~Porfyrakis}
\affiliation{Department of Materials, Oxford University, Oxford OX1 3PH, UK}

\author{Arzhang~Ardavan}
\affiliation{CAESR, Clarendon Laboratory, Dept. of Physics, Oxford University, Oxford OX1 
3PU, UK}

\author{Klaus-Peter Dinse}
\affiliation{Freie Universit\"{a}t Berlin, Fachbereich Physik, Arnimallee 14
14195 Berlin, Germany}

\author{G.~Andrew~D.~Briggs}
\affiliation{Department of Materials, Oxford University, Oxford OX1 3PH, UK}

\date{\today}

\begin{abstract}

We present an electron paramagnetic resonance (EPR) study of
\erscnC~fullerene in which there are two Er$^{3+}$ sites
corresponding to two different configurations of the ErSc$_{2}$N
cluster inside the \ceighty~cage. For each configuration, the EPR
spectrum is characterized by a strong anisotropy of the
\emph{g}-factors (\emph{g}$_{x,y}$ = 2.9, \emph{g}$_{z}$ = 13.0 and
\emph{g}$_{x,y}$ = 5.3, \emph{g}$_{z}$ = 10.9). Illumination within
the cage absorption range ($\lambda<600$~nm) induces a
rearrangement of the ErSc$_{2}$N cluster inside the cage. We
follow the temporal dependence of this rearrangement phenomenologically under
various conditions. 
\end{abstract}

\maketitle

\section{INTRODUCTION}

Rare-earth (RE) doped fullerenes have been widely studied because of
their interesting structural, magnetic and optical
properties~\cite{RPP.63.843,Smirnova_CPL_2008,CPL.343.229}. Many
studies investigate optical properties and are devoted to the
determination of the electronic structure of the RE ions inside the
cage~\cite{RPP.63.843,Dunsch_JPCS_65_309_2004,Krause_JCP2001,PRL_1997}.
However, there is a smaller number of reports about the ground-state
spin characteristics of the doped ions, which can be examined by
electron paramagnetic resonance (EPR)
~\cite{Sanakis_J.Am.Chem.Soc_123_9924_2001,PSSB.243.3037}. The
combination of EPR and optical spectroscopy offers a powerful
approach to understanding the electronic structure and local
symmetries of the ion.

Photoluminescence (PL) has previously been reported from the
Er$^{3+}$~ion in \erscnC~around 1.5~$\mu$m
\cite{CPL.343.229,JCP.127.194504,TiwariPSSB2008}, revealing that the
ErSc$_{2}$N cluster inside the \ceighty~cage occupies two dominant
spatial configurations, corroborating earlier X-ray crystallographic
studies \cite{JACS.122.12220}. At low temperatures, one of the
configurations is favoured over the other, and illumination with
visible light was found to switch the \erscn~molecular cluster
inside the cage from one configuration to the other~\cite{John.sub}.

In this article we report a detailed EPR investigation of \erscnC,
including a study on the switching of two configurations of the
\erscn~cluster within the cage.
\section{Materials and Methods}
\erscnC~was purchased from Luna Innovations with an approximate
purity of 60\%. Impurities included other erbium doped
metallofullerenes. It was further purified to $>$95\% by high
performance liquid chromatography (HPLC). Solutions were prepared in
toluene ($>$99\%, Fisher Scientific) and o-terphenyl (99\% Aldrich) with a minimum concentration of 
$3\times10^{-4}$ M
(approximately $10^{15}$~\erscnC~molecules). The solution was
freeze-pumped in liquid nitrogen for several cycles to deoxygenate
the solvent and was sealed under vacuum in a 4~mm quartz EPR tube.
CW EPR spectra were recorded in the temperature range 5~K to 30~K
using an X-band Bruker EMX Micro EPR spectrometer equipped with a
rectangular TE$_{102}$ cavity (quality factor of about 12000) with
an optical window and an Oxford ESR900 helium flow cryostat. The
sample was illuminated using a range of sources: a Nd:YAG laser
($\lambda=532$~nm); a tunable infra-red laser
($\lambda=1420-1520$~nm); and light emitting diodes (LEDs) of 10 mW
power emitting at 400 nm, 470 nm and 525 nm. Throughout, the
microwave frequency was 9.4~GHz. The magnetic field was swept in the
range 0 to 400~mT, modulated with an amplitude of 0.3~mT at a
frequency of 100~kHz.

\section{EPR spectra and the hyperfine structure}
Assuming a completely ionic model, the electronic configuration of
\erscnC~can be written as
Er$^{3+}$(Sc$^{3+}$)$_{2}$N$^{3-}$@C$_{80}^{6-}$
\cite{Krause_JCP2001}. Sc$^{3+}$, N$^{3-}$~and C$_{80}^{6-}$~have no
unpaired electrons, whereas the Er$^{3+}$~ion, with
4$f^{11}$~electrons in its valence shell, is presumably the only
EPR--active species inside \erscnC. Erbium has six natural isotopes,
$^{162}$Er, $^{164}$Er, $^{166}$Er, $^{167}$Er, $^{168}$Er and
$^{170}$Er. Only $^{167}$Er, with a natural abundance of 23\%, has
an effective nuclear spin $I=7/2$. The other isotopes have $I=0$.
EPR of rare-earth (RE) ions in a weak crystal-field is normally
observed only at low temperatures, owing to fast spin-lattice
relaxation~\cite{abragam70}. At 5~K, only the lowest doublet is
expected to be populated giving rise to the resonance spectrum with
an effective spin $S_{\mathrm{eff}}=1/2$.

\begin{figure}[t]
\includegraphics[width=8cm]{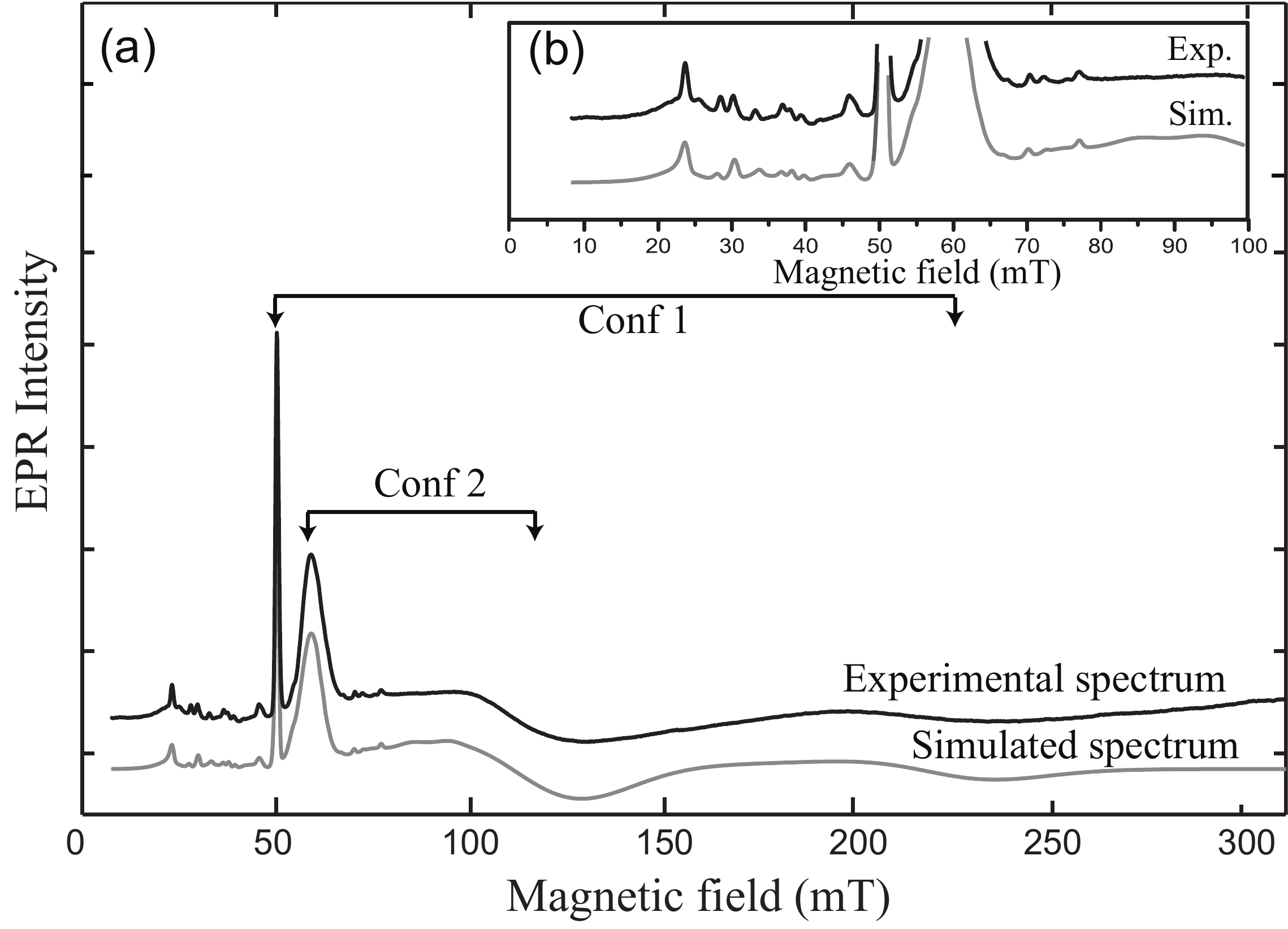}\\
\caption{(a) X-band CW EPR spectrum of \erscnC~in a frozen toluene
solution at 5~K and EasySpin simulated spectrum.~(b)~Expanded
spectra of the experimental and simulated curves in the low-field
region showing the resolved hyperfine splitting.} \label{fig1}
\end{figure}

The EPR spectrum of \erscnC~in toluene was only observed at
temperatures below 30~K. The spectrum of \erscnC~in toluene at 5~K
is shown in \fig{fig1}.  EPR peaks, at 51~mT and 220~mT and at 60~mT
and 110~mT, are attributed to two non-equivalent Er$^{3+}$~ions
inside the cage, corresponding to two orientation of the internal
cluster within the molecule~\cite{John.sub}. Additional hyperfine
structure is also observed around the narrow low-field lines.
EasySpin software~\cite{stoll.178.42} was used to simulate the
experimental spectrum, with an effective spin-Hamiltonian:
\begin{equation}
H_{\mathrm{eff}}=\mu_{\mathrm{B}}gB.S_{\mathrm{eff}}+\mu_{n}g_nB.I+S_{\mathrm{eff}}.A.I+I.Q.I
\label{eq1}
\end{equation}
where the four terms correspond to the electron- and nuclear spin
Zeeman interaction, the hyperfine interaction and the
nuclear--quadrupolar interaction respectively.

\begin{table}
\caption{\label{tab1}Simulated principal values of \emph{g} and
\emph{A}~(in MHz) of Er$^{3+}$~in \erscnC~for \sitea~and \siteb. The
$A_{x}$ and $A_{y}$ values and the components of \emph{Q}~(in MHz)
are only reported for Er$^{3+}$ in \sitea; for Er$^{3+}$ in \siteb,
the effect of these values remains unresolved.}
\begin{ruledtabular} \begin{tabular}{c|ccc|cc}
 & \multicolumn{3}{c}{Er$^{3+}$, \sitea} & \multicolumn{2}{c}{Er$^{3+}$, \siteb}\tabularnewline
\hline Axis & \emph{g} & \emph{A} & \emph{Q} & \emph{g} &
\emph{A}\tabularnewline \hline
\emph{x} & 2.9 (0.3)  & 380 (50) & 10
(5) & 5.3 (0.3) & NA\tabularnewline
\emph{y} & 2.9 (0.3) & 380 (50)
& 130 (5) & 5.3 (0.3) & NA\tabularnewline
\emph{z} & 13.0 (0.1) &
1380 (10) & 10 (2) & 10.9 (0.1) & 1590 (20)\tabularnewline
\end{tabular}\end{ruledtabular}
\end{table}

The simulated spectrum, as shown in \fig{fig1}, reproduces the experimental one accurately. The relative intensities of the EPR peaks are simulated using the natural Er isotopic abundance. The simulation parameters are given in \tab{tab1}. The \emph{g}-, \emph{A}- and \emph{Q}- tensors are characterised by one large and two small components, revealing the large anisotropy in the system. We extract a set of values for Er$^{3+}$ in each of the two configurations in \erscnC. Owing to the narrow linewidth (0.8 mT) of the transition at 51~mT for \sitea, further splitting caused by the nuclear--quadrupolar interaction \emph{Q} was observed. 
However, for the other configurations, \siteb, the values of $A_{x}$ and $A_{y}$ could not be determined accurately because of the large linewidths and the low intensities of the corresponding peaks of \siteb. Similarly, the effect of the quadrupolar interaction is smeared out over the respective hyperfine lines. 

The results of \tab{tab1} seem to suggest an axial symmetry for the
tensors. However, from previous X-ray diffraction measurements the
planar \erscn~cluster has been found to possess a C$_{2\textit{v}}$
(or orthorhombic) symmetry \cite{JACS.122.12220}. In addition,  the
Er$^{3+}$ ion sits probably in an even lower symmetry site if one
considers the neighbouring carbon atoms of the C$_{80}$ cage, which
are located 2.5 to 3~\AA~from the ion \cite{JACS.122.12220}. Since
the uncertainties obtained from our fitting of the \emph{g}-,
\emph{A}- and \emph{Q}- tensors are relatively large, we then
conclude that these tensors have in fact no particular symmetry.

The values of the ratios $A_{x}/g_{x}$, $A_{y}/g_{y}$ and
$A_{z}/g_{z}$ for \sitea~are not equal. This means that the crystal
field admixes significant amounts of  different \{J, m$_{J}$\}
states to the ground state \cite{abragam70}. The presence of such
admixtures has also been shown in the closely related
\ernC~fullerene from photoluminescence spectroscopy and
crystal-field modelling \cite{Jones_thesis}. It is interesting that
the results obtained by Guillot-Noel \textit{et al.}\ for Er:YSO
crystals where the Er sits in C$_{1}$ symmetry sites, give similar
values for the hyperfine parameters and thus, significant mixing
between \{J, m$_{J}$\} states \cite{ogn2006}.

As another consequence of the low symmetry of the Er$^{3+}$ in
\erscnC, the \emph{Q} tensor axes are not coincident with respect to
the \emph{g} and \emph{A} tensors \cite{pilbrow_low-symmetry_1980}.
The high $Q_y$ value is about the same order of magnitude as the
\emph{x} and \emph{y} components of the hyperfine tensor, thus
making nuclear electric quadrupole transitions observable in the EPR
spectrum as additional satellite peaks near the intense 51~mT peak.

It is extremely difficult to obtain more structural information from
all the tensor values as the EPR spectrum is a powder spectrum,
further complicated by the presence of two configurations whose
linewidths are very different. Further information could be obtained by
aligning these molecules, for example with a high magnetic field (some
magnetic orientation has been shown on \ernC~in CS$_{2}$ with a
field of 19 T \cite{Jones_thesis,jones2}) and study the EPR spectra at
various sample rotation angle.

\section{Line broadening}
At low temperatures, the EPR linewidth $\delta\omega$ is affected by
inhomogeneities in the sample such as the \emph{g}- and
\emph{A}-strains, unresolved hyperfine couplings, and also by
homogeneous effects which limit the linewidths to 1/$T_{2}$, where
$T_{2}$ is the spin--spin relaxation time~\cite{PR.70.460}. As the
temperature rises, the spin-lattice relaxation time ($T_{1}$)
shortens, limiting $T_2$ and leading to broadening of EPR peaks.

\begin{figure}[b]
\includegraphics[width=8.5cm]{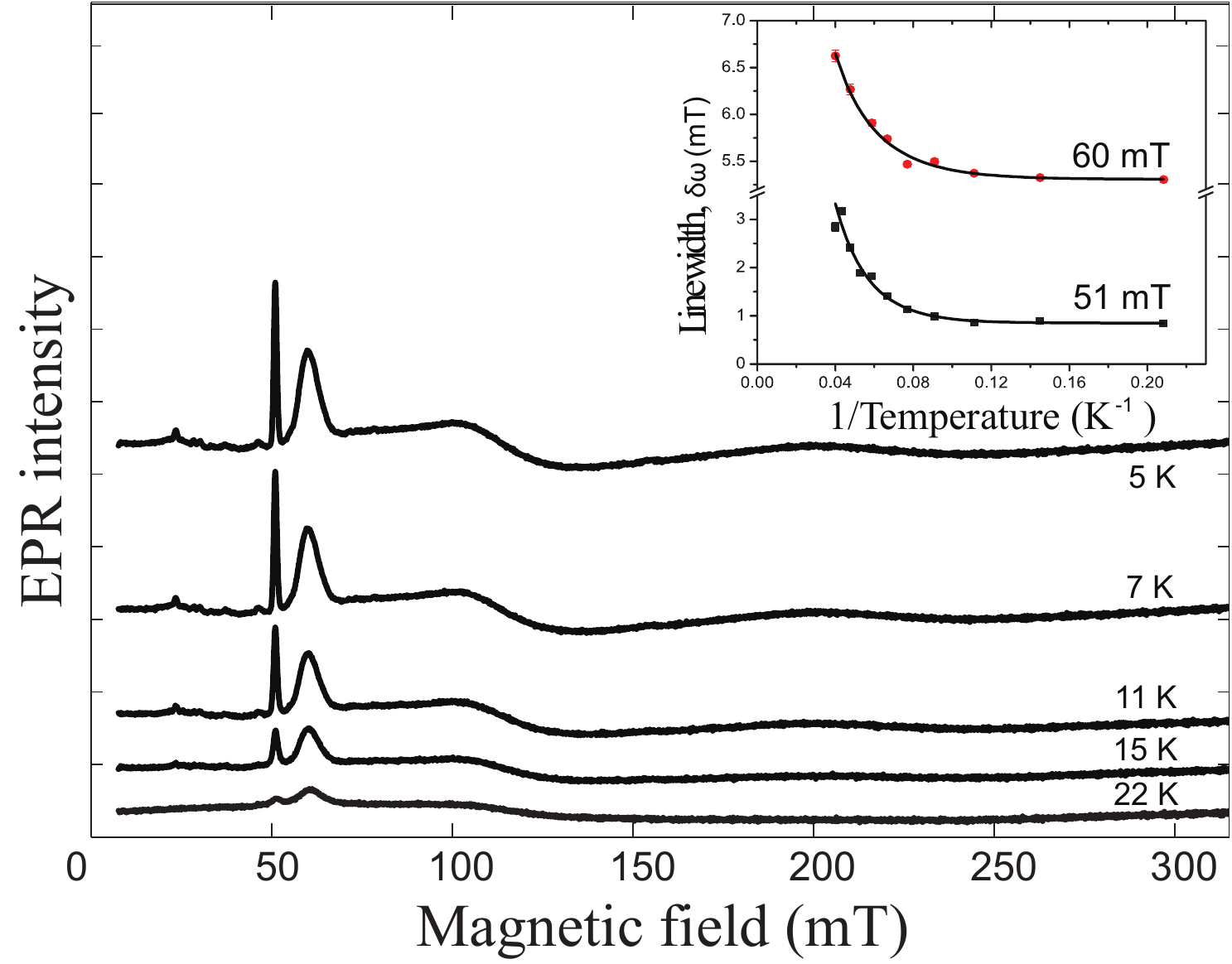}\\
\caption{EPR spectra of \erscnC~at different temperatures. The inset
shows the temperature dependence of linewidths of two most intense
peaks at 51~mT and 60~mT observed in the EPR spectrum at 5~K. The
symbols and the solid curves indicate the experimental data and the
fitted curve, respectively. The activation energy calculated from
fitting the evolution of $\delta\omega$ as a function of temperature
support the assignment of these peaks to \sitea~and
\siteb~respectively.} \label{fig2}
\end{figure}

\Fig{fig2} shows EPR spectra of \erscnC~at different temperatures.
The inset shows the evolution of linewidths of the peaks at 51~mT
and 60~mT as a function of 1/T, which fit well to the general
expression of the form:
\begin{equation}
\delta\omega=\frac{\mathrm{S}}{[\exp(\frac{\Delta_{E}}{k_{B}T})-1]}+\mathrm{Constant}\label{eq3}
\end{equation}
where $k_{B}$~is the Boltzmann constant, S is the magnitude of the
linewidth normalized by the phonon occupation number and
$\Delta_{E}$~is the energy of an excited level. This dependence
reflects that the spin-lattice relaxation is an Orbach mechanism,
i.e.\ a two phonon $T_{1}$~process where the relaxation takes place
via an excited state~\cite{prc.a264.458}.

The energy $\Delta_{E}$, obtained from the fitting parameters of \eqn{eq3} for the two peaks at 51~mT and 60~mT, is found to be 36(1)~cm$^{-1}$ and 25(4)~cm$^{-1}$, respectively. PL spectroscopy has previously revealed the crystal-field energy levels of the Er$^{3+}$ ion in two different configurations, corresponding to two different orientations of the ErSc$_{2}$N cluster within the \ceighty~cage \cite{JCP.127.194504}. The energy difference between the lowest and second lowest crystal-field levels of the ground state \gs~of Er$^{3+}$~ion is 37~cm$^{-1}$~for \sitea~and 28~cm$^{-1}$~for \siteb. $\Delta_{E}$, deduced from the EPR studies, is in good agreement with the energy differences obtained in PL. Thus, EPR peaks at 51~mT and 60~mT are assigned to the optically observed \erscn~configurations, \sitea~and \siteb, respectively.

\section{High-frequency EPR}
\begin{figure}[t]
\includegraphics[width=8cm]{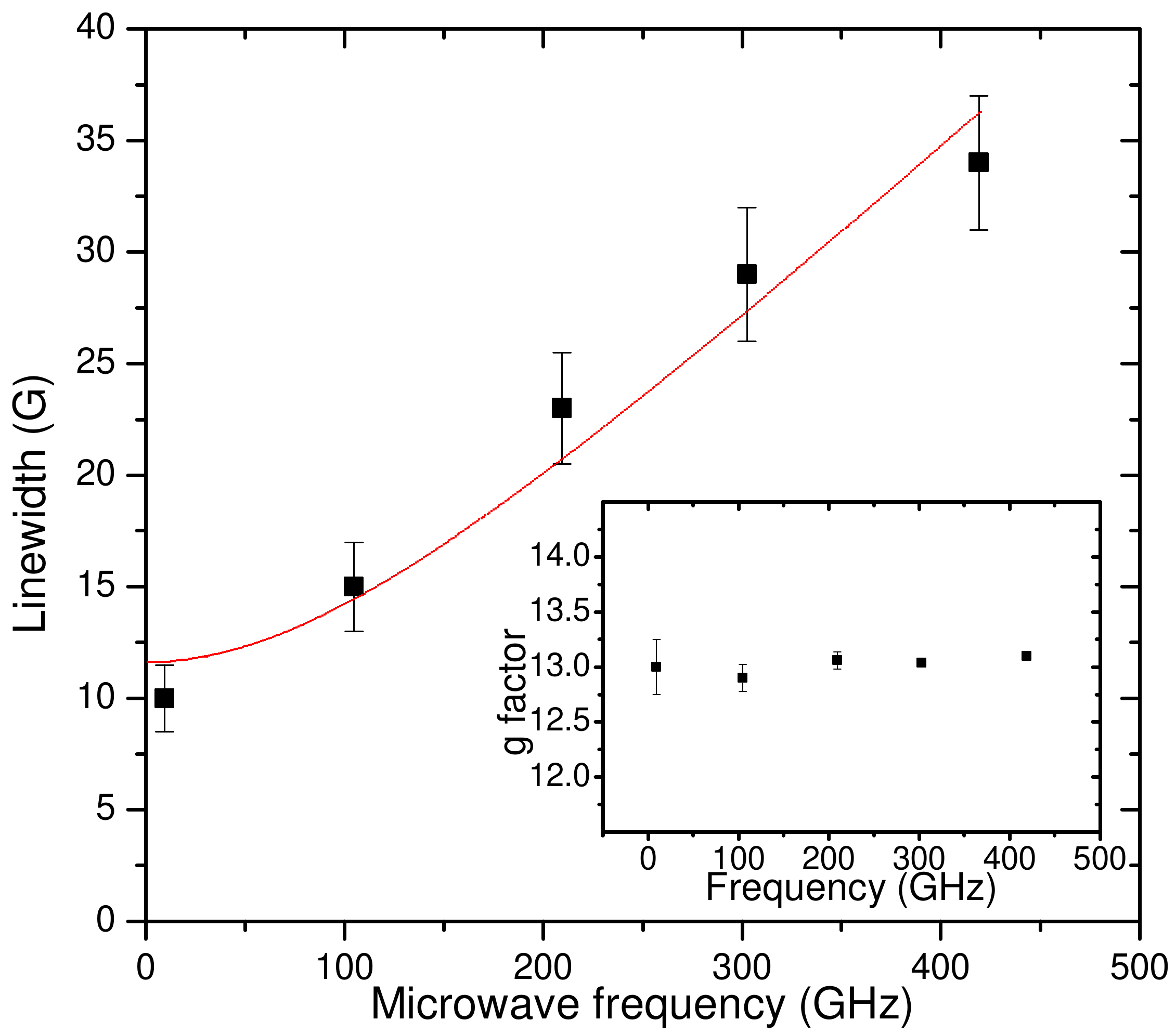}\\
\caption{Linewidth and g-factor (inset) of $g_{z}$ feature of
\sitea~as a function of microwave frequency. The solid line is a fit to a combination of a frequency-independent homogeneous component to the linewidth (12~G), and a g-strain component which is linear with frequency (82~mG/GHz). 
} \label{hfepr}
\end{figure}

High frequency CW-EPR of \erscnC~was performed at the National High Magnetic Field Lab (NHMFL), Tallahassee using their home-built 100--600~GHz homodyne spectrometer~\cite{SchneiderMuntau2004643}. Figure~\ref{hfepr} shows the linewidth dependence of the narrowest spectral feature, the $g_{z}$ feature of \sitea, as a function of microwave frequency (and thus magnetic field). The data are consistent with the linewidth at X-band (10~GHz) being homogeneously broadened (on which assumption the analysis above relied), while at higher frequency $g$-strain across the sample causes a linear increase in linewidth. The g-strain which can be extracted is  small: $\approx 0.15\%$ indicating its value shows a very weak dependence on local environment.

\section{EPR spectra upon illumination}

\subsection{Wavelength dependence}

\begin{figure}[h]
\includegraphics[width=8.5cm]{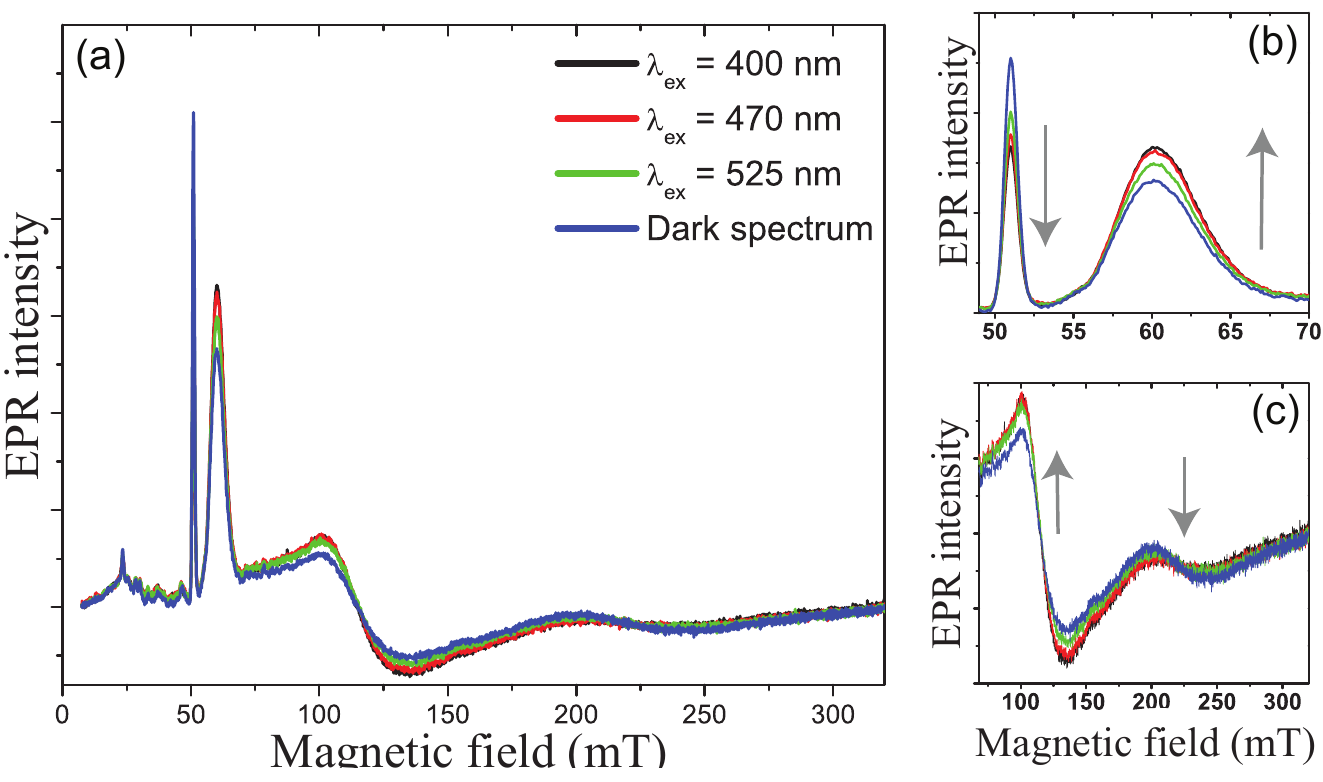}\\
 \caption{(a) Influence of illumination at 400 nm, 470 nm and 525 nm on the EPR signal of Er$^{3+}
$~ion in \erscnC~ in toluene at 5~K. The figures (b) and (c) show changes in the peak intensities at 
51~mT and 60~mT, and 220~mT and 110~mT. For a comparison, the EPR spectrum of a non-
illuminated sample is also shown.}
\label{fig3}
\end{figure}

In this section we discuss the influence of optical illumination on the EPR spectrum of \erscnC. For each excitation wavelength, the sample was illuminated for 20~mins at 5~K before the EPR spectrum was recorded. Between measurements at different wavelengths, the sample was heated above 35~K and then cooled down to 5~K. This annealing process provided the same initial conditions before each new period of illumination.

Initially, we illuminated the Er$^{3+}$ ions to excite transitions within the crystal-field levels directly, using excitation wavelengths of 1496 nm and 1499 nm~\cite{JCP.127.194504} for \sitea~and \siteb~respectively. These wavelengths are too long to excite the cage and had no effect on the EPR spectrum.

We then excited \erscnC~at different wavelengths well within the cage absorption range using 400~nm, 470~nm and 525~nm LEDs. The EPR spectra of \erscnC~under LED illumination are shown in \Fig{fig3}. For comparison, the EPR spectrum without illumination is also shown on the same graph. The spectra reveal that the intensities of the EPR peaks corresponding to the two \erscn~configurations ions vary in opposite directions (indicated bt the arrows), allowing the assignment of the set of peaks at 51~mT and 220~mT to \sitea~and the set of 60~mT and 110~mT peaks to the other \siteb~\cite{John.sub}. The line shape and spin-Hamiltonian parameters remain unchanged on illumination, and no evolution of any new spectral features is observed.

The change in the EPR intensity for each configuration indicates that the ErSc$_{2}$N cluster rearranges within the \ceighty~cage upon illumination. Exciting \erscnC~with visible light appears to preferentially drive the switching one configuration to the other. As the direct excitation of Er$^{3+}$ ion has no influence on the relative proportion of the two configurations,  we can conclude that the configurational changes are mediated via the cage. Subsequently, we used a green laser operating at 532~nm as an excitation source. The resulting change in the EPR spectrum was consistent with the effect of illumination using LEDs. Probably owing to the higher excitation power (26~mW, compared to 10~mW for the LEDs), illumination with the 532~nm laser induced a 10\%~larger change in the EPR intensities as compared to the illumination with the 525~nm LED in the given time span of 20~mins.

\subsection{Temperature dependence}

We studied the effect on the EPR spectrum of laser illumination at 532~nm at various temperatures, $T$, between 5~K and 15~K. The integrated area of the spectrum depends on many factors, including transition moment, temperature, microwave power and relaxation times. We combine all such terms into a constant of proportionality $\kappa$ such that the intensity of a feature attributed to \sitei, $I_{i} =\kappa_{i} \, n_{i}$, where $n_{i}$ is the number of spins in \sitei.

\begin{figure}[t]
 \includegraphics[width=8.5cm]{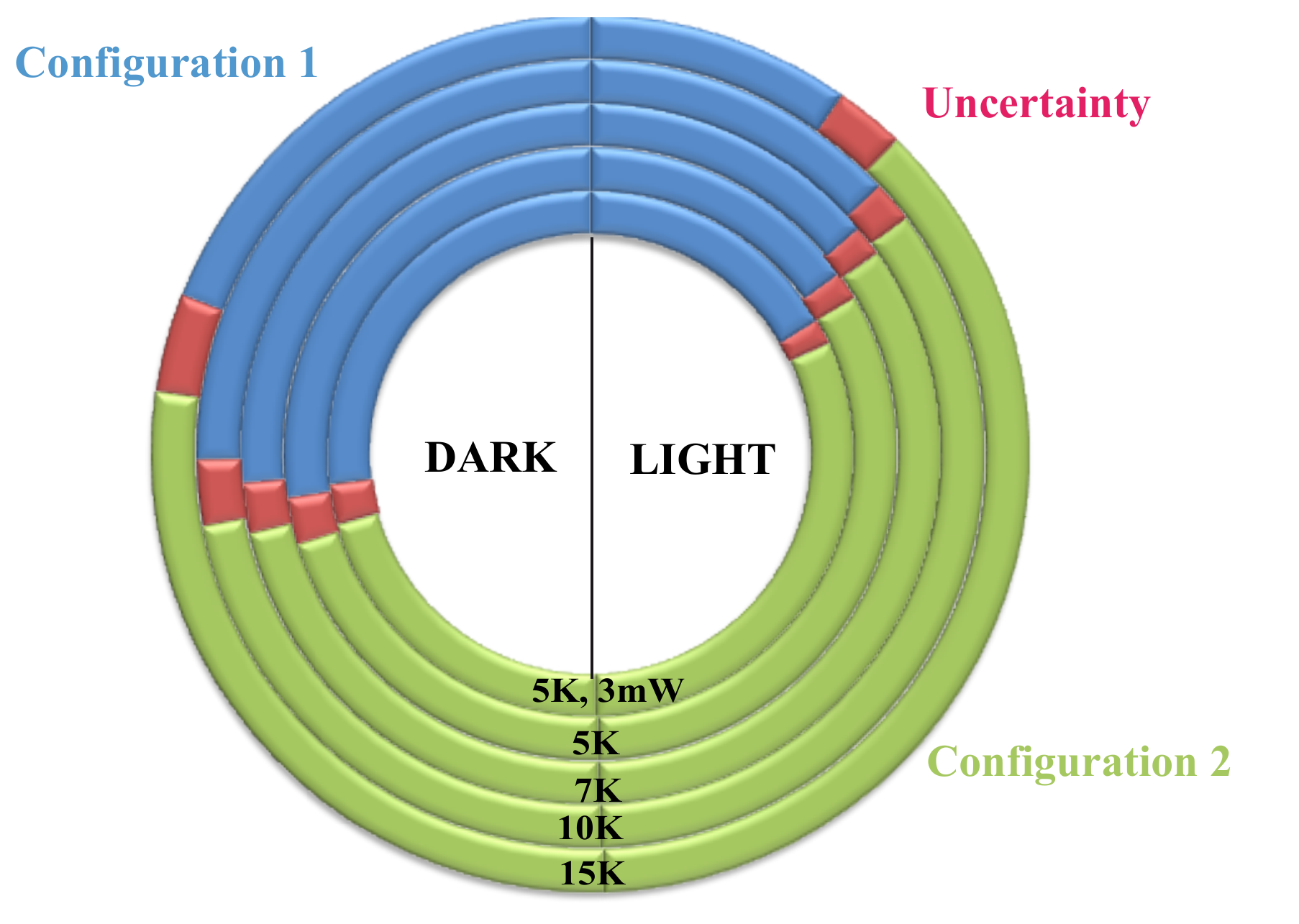}\\
\caption{Populations of \sitea~and \siteb~from the ESR peaks of \erscnC~in o-terphenyl for various experimental conditions. The exictation laser was set to 532~nm and 20~mW power, except for one measurement as indicated in the figure. The sample was annealed at room temperature between each measurement.}
\label{donut}
\end{figure}

Assuming that $\kappa_i$ for a particular configuration is independent of the illumination, the change in the integrated area under the resonance peak provides a measure of the change in the configuration populations. We examined the resonance peaks at 51~mT and 60~mT corresponding to \sitea~and \siteb~respectively. Figure \ref{donut} shows the change in the populations for \sitea~and \siteb~at various temperatures with and without illumination. The populations were obtained by a fitting procedure assuming that the total number of spin is constant and that only two configurations are present. The latter assumption has been verified experimentally (see the population switching dynamics section).

Previously, it has been shown that at 5~K a swing in configuration population by as much as $\approx27\%$ \siteb is possible~\cite{John.sub} upon optical illumination. In order to study the effect on switching of various parameters, such as temperature, it is important to control the molecular environment carefully. The use of toluene as a solvent produces a metastable glass upon freezing which easily cracks and form a polycrystal. In such a polycrystal, the various orientations of the grains induce strains that probably impact on the local environment of the fullerene, and hence the Er$^{3+}$ ions. We therefore 
used o-terphenyl as a solvent as it is known to produce stable glasses, and therefore a more reproducible environment for the \erscn. In this case, the sample needs to be first heated to about 330~K and then quickly frozen in liquid nitrogen.  
We observe in figure \ref{donut} that the \siteb~is predominant at higher temperatures. Moreover, a weaker laser illumination results in less switching of \sitea~to \siteb. To summarise, the analysis of the populations before and after illumination shows that higher temperatures and illumination both favour \siteb .

\section{Population switching dynamics upon illumination}

\subsection{Evidence for the switching between only two configurations ErSc$_{2}$N cluster}
Here we report the experimental dynamics of the population switching between two configurations of ErSc$_{2}$N cluster within the \ceighty~cage upon illumination.

The magnetic field range was selected to include the two sharp peaks at 51~mT and 60~mT. The sample was diluted in o-teprhenyl as it formed a better glass than toluene. After being kept in dark for about 5~min, it was illuminated with a laser at 532~nm and 20~mW power for 48~min. The changes in EPR peaks at 51~mT and 60~mT were recorded simultaneously. 

As the laser is turned on, the intensity of the peak at 51~mT, corresponding to \sitea, decreases and the one for the peak at 60~mT, corresponding to \siteb, increases.

\begin{figure*}
\includegraphics[width=16cm]{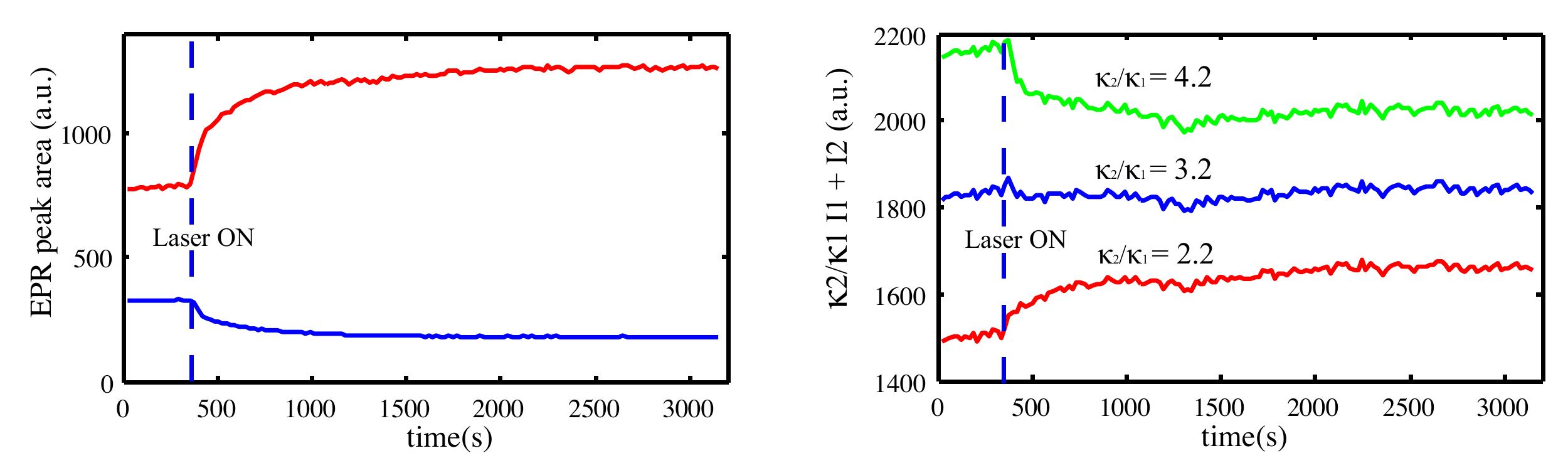}\\
\caption{(a) Temporal evolution of the integrated area under the EPR
peaks at 51~mT and 60~mT before and during illumination. Temperature
= 5~K; excitation wavelength $\lambda=532$~nm; power  = 20~mW. (b)
Plot of $\kappa_{2}/\kappa_{1}*I_{1}+I_{2}$) versus
time for $\kappa_{2}/\kappa_{1}=$2.2, 3.2 and 4.2.}
\label{fig7}
\end{figure*}

The integrated area ($I_{i}$) under the two sharp peaks at 51~mT and 60~mT was calculated for each spectrum. \fig{fig7}(a) shows the temporal evolution of the areas ($I_{i}$) before and during illumination.

We assume that the total number of spins $n_1 + n_2 =
\frac{I_1}{\kappa_1}+ \frac{I_2}{\kappa_2}$ is constant throughout
illumination. We test this assumption and derive the ratio of the intensity factor
$\kappa_i$, by plotting ($\frac{\kappa_{2}}{\kappa_{1}}I_{1}+I_{2}$) as function of
time for different ratios of $\kappa_{2}/\kappa_{1}$. Through a fitting procedure we find that for $\kappa_{2}/\kappa_{1}=3.2\pm0.1$, the function
($\frac{\kappa_{2}}{\kappa_{1}}I_{1}+I_{2}$), which is proportional to the total number of spins in \sitea~and \siteb, is constant. This is illustrated in \fig{fig7}(b) for a range of $\kappa_{2}/\kappa_{1}= 2.2, 3.2, 4.2$. This value can be approximately understood by considering the different total spectral widths ($g_{\perp}$ to $g_{\parallel}$) of the two configurations.

As $t\rightarrow\infty$, the area under the two peaks corresponding
to each configuration, $I_{1}(\infty)$~and $I_{2}(\infty)$,
reaches a steady regime under a constant illumination (see \Fig{fig7}(a)).
From \fig{fig7}(a), the ratio $I_{1}(\infty)/I_{2}(\infty)$
can be evaluated. As we know that: \begin{equation}
\frac{I_{1}(\infty)}{I_{2}(\infty)}=\frac{n_{1}(\infty)\kappa_{1}}{n_{2}(\infty)\kappa_{2}}\end
{equation}
 and the ratio $\kappa_{2}/\kappa_{1}=3.2$, we can deduce the ratio $n_{1}(\infty)/n_{2}(\infty)$.
At 5~K and 20~mW, the ratio is:
\begin{equation}
\frac{n_{1}(\infty)}{n_{2}(\infty)}=0.45(2)
\end{equation}
This ratio is in agreement with that reported previously~\cite{John.sub}. We note that this shows the light-induced switching observed is not due to sample heating, which would reduce the intensity of both peaks, as shown above. 

\subsection{Effects of temperature and power on the temporal evolution}
\begin{figure*}
\includegraphics[width=16cm]{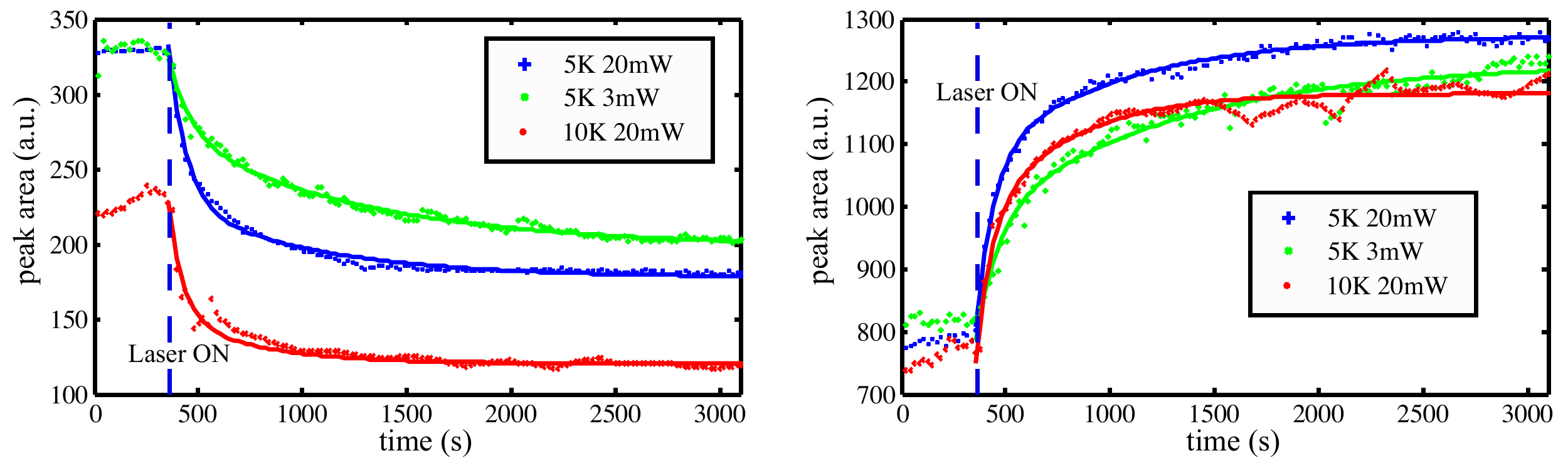}\\
\caption{(a) Time response of the integrated area under the peak at 51~mT
before and during illumination at T = 5~K and 10~K with an excitation
wavelength $\lambda=532$~nm and power of 20~mW and 3~mW. (b) Time
response of the integrated area under the peak at 60~mT before and
under illumination at T = 5~K and 10~K with an excitation wavelength
$\lambda=532$~nm and a power of 20~mW. The laser was turned on after 400s. Smooth curves show biexponential fits with the parameters from the model.}
\label{fig8}.
\end{figure*}

In order to gain further insight on the switching dynamics, we performed similar time scan measurements at a lower power and higher temperatures. The temporal evolution of
the integrated area under various conditions for each configuration is shown in \fig{fig8}. We confirmed that for most conditons, the ratio $\kappa_{2}/\kappa_{1}$ for 3~mW is unchanged at $3.2\pm0.1$. Only at 10~K this ratio is slighlty higher, with $\kappa_{2}/\kappa_{1} = 3.8\pm0.1$. Thus, we can directly translate the changes in EPR peak areas to the populations dynamics of the configurations.

The data are well fit to biexponential decay functions of time, with a slow decay time ($T_s$) and a fast decay time ($T_f$). We consider the average decay time $T_{avg}$ = ($A_sT_s$+$A_fT_f$), where $A_s$ and $A_f$ are the normalised amplitudes of the two exponential decay components. The results are presented in \tab{tab3}. Reducing laser power from 20 to 3~mW has no significant effect on the extracted ratios of $n_1(0)/n_2(0)$ and $n_1(\infty)/n_2(\infty)$, as expected from Figure 5, however the average decay time $T_{avg}$ is much slower. 

In contrast, changing the temperature from 5 to 10~K involves a change in both the initial and final population ratios (as observed above). However it does not affect $T_{avg}$.  Thus, the temporal evolution of the switching process is dominated by the kinetic effect of laser power, rather than a thermodynamic effect.

Simple kinetic models involving different excitation and relaxation rates can reproduce the basic form of this biexponential behaviour, however a quantitative comparison would require a detailed understanding of the energy transfer process between the carbon cage and the Er$^{3+}$~ion . Further optical studies on \erscn\ and related TNT fullerenes will help to elucidate this transfer process and the energy levels involved.

\begin{table}
\begin{ruledtabular} \begin{tabular}{cccccccc}
T (K) & Power (mW) & $\frac{n_{1}(0)}{n_{2}(0)}$ & $\frac{n_{1}(\infty)}{n_{2}(\infty)}$ & $T_{avg}$ (s) \tabularnewline
\hline
5 & 20 & 1.33(0.02) & 0.45(0.02) & 302(60)
\tabularnewline
\hline
5 & 3 & 1.28(0.02) & 0.50(0.02) & 613(100) 
\tabularnewline
\hline
10 & 20 & 1.12(0.02) & 0.37(0.02) & 309(60)
\end{tabular}\end{ruledtabular}
\caption{\label{tab3} Average decay times deduced from the fitting parameters of the
bi-exponential temporal evolution of the average integrated area as a function of temperature and laser power.}
\end{table}

\subsection{Relaxation after illumination}
When the laser is turned off, the peak intensities evolve in the
opposite direction in a monoexponential way. This is consistent with
a direct relaxation between \siteb~and \sitea. This relaxation is
very slow compared to the excitation timescales: after 12 hours at
15~K, the configurations relax only half way to equilibrium. Thus at
low temperatures, the configuration populations are very stable,
while they recover to equilibrium very rapidly if the sample is
heated above 35~K.

\section{Conclusions}
We have investigated the behaviour of two dominant configurations,
\sitea~and \siteb, of the Er$^{3+}$~ion in \erscnC~using continuous
wave EPR. The experimental spectra can be simulated to extract a spin Hamiltonian with
anisotropic $g$- and $A$- and $Q$- tensors for each cluster configuration, to the extent allowed by the EPR linewidths. The results suggest that
the erbium ion  sits in a low symmetry site, at least for
\sitea. The two configurations are switchable through optical
excitation of the \ceighty~cage. The switching dynamics can be
described phenomenologically by biexponential functions of time.
The average decay time becomes slower at lower laser power but does not change with temperature.
The recovery
to equilibrium populations is very slow at low temperatures, but the
equilibrium populations can be quickly recovered by annealing the
sample above 35~K. This thermo-optical hysteresis makes \erscnC~a
potential candidate for reversible switches and memory
elements~\cite{zhang:158301,ThorstenHugel05102002,AM.14.293,Iwamoto1991,Zhao_CPL_1999,Correa_Poly2006,berkdemir-2005-254}.

\section{Acknowledgements}
Authors would like to thank Marshall Stoneham, William Hayes and
Alexei Tyryshkin for valuable discussions.  AT would like to
thankfully acknowledge Dr. A. Tripathi and E. Gauger for their
technical support. This research is supported by the EPSRC through
the QIP IRC www.qipirc.org (No. GR/ S82176/01), and the Oxford
Centre of Advanced Electron Spin Resonance (No. EP/D048559/1). RR is supported by 
the Marie Curie QIPEST program and the EPSRC. AT is
supported by Felix Scholarship. GD is supported by QIP IRC, Oxford.
JJLM  and AA are supported by the Royal Society. GADB is supported by
EPSRC (GR/S15808/01). Unpurified TNT sample was supplied by Luna
Innovations, Blacksburg, Virginia, USA.

\bibliography{biblio}

\end{document}